\documentclass[aip,pop,groupedaddress,preprint]{revtex4-1}
\usepackage{physics}
\usepackage{amsmath}
\usepackage{graphicx}
\usepackage{subfigure}
\usepackage{hyperref}
\begin{document}
\title{An improved approach to derive the kappa distribution in polytropic plasmas} 
\author{Ran Guo}
\thanks{Author to whom correspondence should be addressed}
\email{rguo@cauc.edu.cn}
\affiliation{Department of Physics, College of Science, Civil Aviation University of China, Tianjin 300300, China}
\begin{abstract}
    This paper studies sufficient conditions for deriving the kappa distribution in polytropic plasmas by an improved method compared with the previous work [R. Guo, Phys. Plasmas \textbf{27}, 122104 (2020)]. 
    We find that the polytropic equation of state can lead to the kappa distribution without any other assumptions in one dimension.
    In higher dimensions, an extra assumption that the global distribution must only depend on the energy is still needed.
    In addition, the self-consistence of the electrostatic potential is investigated in such plasmas. 
    The study implies that in single-component plasmas, the potential is constrained by the Lane-Emden equation.
    However, in two-component plasmas, any arbitrary potential could exist with a suitable background ion density. 
    Finally, we discuss the connections between the kappa distribution and the polytrope.
\end{abstract}
\maketitle

\textit{Introduction.} Space plasmas are always far from thermal equilibrium.
As a result, such plasmas cannot be described by the Maxwellian distribution very well.
A more suitable model is the kappa distribution supported by several studies in various plasmas, such as solar wind, \cite{Maksimovic1997,Pierrard2016}
solar corona, \cite{Vocks2008,Cranmer2014} and planetary magnetosphere. \cite{Dialynas2009}
A common type of the 3-dimensional (3-D) kappa velocity distribution may be written as, \cite{Livadiotis2010a,Pierrard2016}
\begin{equation*}
    f_\kappa(v)  \propto \left( 1 + \frac{1}{\kappa-\frac{3}{2}}\frac{m v^2}{2k_B T} \right)^{-\kappa-1},
\end{equation*}
where $T$ is the temperature and $\kappa$ is the parameter which measures the distance away from the Maxwellian distribution.
The kappa distribution has also been applied to analyzing the physical properties of the plasmas with suprathermal particles. \cite{Lazar2016,Lazar2010,Tribeche2012,Du2013}
Nevertheless, the physical explanation of the generation of the kappa distribution is still under discussion.
Hasegawa et al. prove that the plasma in a suprathermal radiation field follows the kappa distribution. \cite{Hasegawa1985}
Vocks and Mann show that the resonant wave-particle interaction with whistler waves leads to the generation of the suprathermal particles. \cite{Vocks2003}
This is also supported by another particle-in-cell simulation. \cite{Tao2014}
Yoon finds that the kappa distribution can be generated due to the Langmuir turbulence. \cite{Yoon2014}
Leubner \cite{Leubner2004,Leubner2005} and Livadiotis \cite{Livadiotis2009} suggest that the kappa distribution can be explained by the nonextensive statistical mechanics. 

Besides these above physical mechanisms, the kappa distribution can also be generated in an inhomogeneous plasma in which the polytropic equation of state (EOS) holds.
Many theoretical works reveal that the polytropic behavior and the kappa distribution are deeply related to each other.
On the one hand, one can derive the polytropic EOS directly from the kappa distribution. \cite{Scudder1992,MeyerVernet1995,Moncuquet2002,Livadiotis2017b}
On the other hand, Livadiotis \cite{Livadiotis2019b} shows that the polytropic EOS also leads to the kappa distribution by the hydrodynamic method.
Our previous work \cite{Guo2020} gives more refined conditions for the generation of the kappa distribution by the kinetic method.
To be specific, it proves that the solution of the steady-state Vlasov-Poisson equations is the kappa distribution in an inhomogeneous isotropic plasma under two assumptions. 
One is the polytropic assumption at the steady state,
\begin{equation}
    T(\vb{r}) \propto n_e(\vb{r})^{\gamma-1},
    \label{assump:previous1}
\end{equation}
where $T(\vb{r})$ is the local temperature, $n_e(\vb{r})$ is the electron density, and $\gamma$ is the polytropic index. 
A number of observations \cite{MeyerVernet1995,Sittler1980,Pang2016} and simulations \cite{Riley2001,Lin2016} support that the polytropic EOS makes a good job of describing the density-temperature relationship in different plasmas.
The other assumption is a constraint on the form of the local velocity distribution.
If we denote $f_e$ as the electron distribution, then the local velocity distribution is defined by $\hat{f}_{e} = f_e/n_e$.
The assumption on $\hat{f}_e$ can be expressed as,
\begin{equation}
    \hat{f}_{e}(\vb{r},\vb{v}) = A[T(\vb{r})] g\left[ \frac{mv^2}{2k_BT(\vb{r})} \right],
    \label{assump:previous2}
\end{equation}
where $A$ is the normalization factor to ensure $\int \hat{f}_e \dd{\vb{v}} =1$, and $g$ is an arbitrary function.
Although one can obtain the kappa distribution successfully, the second assumption \eqref{assump:previous2} restricting the form of the local velocity distribution is still very complicated and thus unsatisfactory.

In this paper, we attempt to improve our previous work \cite{Guo2020} by removing the assumption on the local velocity distribution, namely Eq. \eqref{assump:previous2}.
In one dimension, this can be achieved directly.
However, in higher dimensions, this will be done at the cost of involving a new assumption that global distribution is only a function of energy.
In other words, the polytropic EOS is the unique sufficient condition to derive the kappa distribution only in 1-dimensional (1-D) non-uniform plasma.
In the case of higher dimensions, the assumption about the distribution function is still needed.

\textit{Derivation of the kappa distribution.} We consider the steady state of an isotropic non-uniform Vlasov-Poisson system. 
The ions are set as a static background for simplicity, while the electrons are governed by the steady-state Vlasov-Poisson equations,
\begin{equation}
    \vb{v} \cdot \nabla f_e(\vb{r},\vb{v}) + \frac{e\nabla \varphi(\vb{r})}{m} \cdot  \pdv{f_e(\vb{r},\vb{v})}{\vb{v}} = 0,
    \label{eq:V}
\end{equation}
\begin{equation}
    \nabla^2 \varphi(\vb{r}) = - \frac{e}{\varepsilon_0}[n_i(\vb{r})-n_e(\vb{r})],
    \label{eq:P}
\end{equation}
where $f_e(\vb{r},\vb{v})$ is the electron distribution function, $m$ is the electron mass, $e$ is the elementary charge, $\varphi(\vb{r})$ is the electrostatic potential,
$n_i(\vb{r})$ is the ion density,
and $n_e(\vb{r})$ is the electron density given by 
\begin{equation}
    n_e(\vb{r}) = \int f_e(\vb{r},\vb{v}) \dd{\vb{v}}. 
    \label{def:ne}
\end{equation}
We suppose that the polytropic EOS holds at the steady state,
\begin{equation}
    \frac{T(\vb{r})}{T_0} = \left[ \frac{n_e(\vb{r})}{n_0} \right]^{\gamma-1},
    \label{assump:poly-eos}
\end{equation}
where $\gamma$ is the polytropic index. 
The local temperature $T(\vb{r})$ is defined by,
\begin{equation}
    \frac{d}{2} k_B T(\vb{r}) = \int \frac{1}{2} m v^2 \frac{f_e(\vb{r},\vb{v})}{n_e(\vb{r})} \dd{\vb{v}},
    \label{def:T}
\end{equation}
with the spatial dimension $d = 1,2,3$.  
The notations $n_{0}$ and $T_0$ in Eq. \eqref{assump:poly-eos} are the density and temperature at an arbitrary reference point respectively. 
For convenience, $n_{0}$ is chosen as the average number density, then $T_0$ becomes the corresponding temperature.
The Jeans theorem states that the solution of the steady-state Vlasov equation can be any function of the integrals of the motion. \cite{Binney2008}
As we know, a mechanical system with $d$ degrees of freedom has $(2d-1)$ independent integrals of the motion. \cite{Landau1995}
In the case of $d=1$, there is only one independent integral of the motion and the energy is the most general choice.
Therefore, in one dimension, the steady-state distribution must be a function of energy.
However, this may be not valid in the case of $d>1$, because there is usually more than one independent integral of the motion.
Thus, in higher dimensions, we assume that the electron distribution is just a function of energy.
Therefore, for any dimensions, we have,
\begin{equation}
    f_e(\vb{r},\vb{v}) = f_e(W),
    \label{assump:w-func}
\end{equation}
with the total energy $W = \frac{1}{2}mv^2 + \phi$ and the potential energy $\phi = -e \varphi$.
As a result of Eq. \eqref{assump:w-func}, the fluid velocity of the plasma vanishes everywhere in space,
\begin{equation}
    \int \vb{v} f_e(W) \dd{\vb{v}} = 0,
    \label{eq:zeroflow}
\end{equation}
because $f_e(W)$ is an even function of $\vb{v}$.

If the distribution is a function of the energy, then the electron density \eqref{def:ne} can be rewritten as,
\begin{align}
    n_e         =& \int f_e(W) \dd{\vb{v}} \notag \\ 
                =& \frac{2\pi^\frac{d}{2}}{\Gamma \left(\frac{d}{2} \right)}\int_0^\infty f_e(W) v^{d-1} \dd{v} \notag \\ 
                =& \left(\frac{2\pi}{m}\right)^\frac{d}{2} \frac{1}{\Gamma \left(\frac{d}{2} \right)} \int_\phi^\infty (W-\phi)^{\frac{d}{2}-1} f_e(W) \dd{W}, \notag \\
                =& \left(\frac{2\pi}{m}\right)^\frac{d}{2} D^{-\frac{d}{2}} f_e(\phi)
                \label{dev:ne}
\end{align}
where $D^{-\frac{d}{2}}$ is the Weyl fractional integral operator defined by, \cite{Miller1993}
\begin{equation}
    D^{-\nu} h(t) = \frac{1}{\Gamma(\nu)} \int_t^\infty (\xi-t)^{\nu-1} h(\xi) \dd{\xi}, \quad \Re \nu>0, t>0,
\end{equation}
with an arbitrary function $h(t)$.
It is worth noting that Eq. \eqref{dev:ne} implies $n_e = n_e(\phi)$.
According to the theory of fractional calculus, \cite{Miller1993} $f_e$ could be solved by the fractional derivative,
\begin{equation}
    f_e(W) = \left(\frac{m}{2\pi}\right)^{\frac{d}{2}} D^{\frac{d}{2}} n_e(W),
    \label{eq:fe-calc}
\end{equation}
where
\begin{equation}
    D^{\frac{d}{2}} n_e(W) = \left\{
        \begin{aligned}
            &\left( -\dv{}{\phi}\right)^{\frac{d}{2}} n_e(\phi)\big|_{\phi=W} 
            &\text{when $d$ is even,} \\
            &\left( -\dv{}{\phi}\right)^{\frac{d+1}{2}} D^{-\frac{1}{2}} n_e(\phi) \big|_{\phi=W} 
            &\text{when $d$ is odd.}
        \end{aligned}
    \right.
    \label{eq:fe-calc-dev}
\end{equation}
Therefore, if the analytic form of $n_e(\phi)$ is known, the distribution $f_e$ could be obtained from the above equations.

The analytic solution of $n_e(\phi)$ can be derived from the polytropic EOS \eqref{assump:poly-eos} and zero-flow condition \eqref{eq:zeroflow}.
As a result of the vanishing fluid velocity, the electron pressure is balanced by the electrostatic force,
\begin{equation}
    \nabla p = n_e e \nabla \varphi,
    \label{eq:p-balance}
\end{equation}
where the pressure, a scalar due to the isotropy, is defined by,
\begin{equation}
    p(\vb{r}) = \frac{1}{d}\int mv^2 f_e(\vb{r},\vb{v}) \dd{\vb{v}}.
    \label{def:p}
\end{equation}
The definitions of temperature \eqref{def:T} and pressure \eqref{def:p} ensure that the ideal gas EOS $p = n_e k_B T$ holds.
Substituting the ideal gas EOS into the pressure balance condition \eqref{eq:p-balance}, we derive,
\begin{equation}
    \frac{e\nabla \varphi}{k_B T} = \nabla \ln n_e + \nabla \ln T.
    \label{dev:nphi}
\end{equation}
The correlation between the density and the temperature is assumed to follow the polytropic EOS \eqref{assump:poly-eos}.
Taking the logarithm and then calculating the derivative on both sides of Eq. \eqref{assump:poly-eos}, one obtains
\begin{equation}
    \nabla \ln n_e = \frac{1}{\gamma-1} \nabla \ln T= (-\kappa_0-1) \nabla \ln T,
    \label{eq:dv-poly-eos}
\end{equation}
where we denote the symbol
\begin{equation}
    \kappa_0 = -1 - \frac{1}{\gamma-1},
    \label{eq:kappa-poly}
\end{equation}
to give a familiar form of the kappa distribution.
With Eq. \eqref{eq:dv-poly-eos}, Eq. \eqref{dev:nphi} becomes,
\begin{equation}
    e \nabla \varphi = -\kappa_0 k_B \nabla T.
    \label{eq:du-rela}
\end{equation}
Integrating both sides of the above equation, we get the potential energy,
\begin{equation}
    \phi = -e\varphi = \kappa_0 k_B (T-T_0),
    \label{dev:nphi2}
\end{equation}
in which the zero potential is selected at the point where $T = T_0$.
After substituting Eq. \eqref{assump:poly-eos} into Eq. \eqref{dev:nphi2}, we can find the expression for $n_e(\phi)$,
\begin{equation}
    n_e(\phi) = n_0 \left( 1+ \frac{\phi}{\kappa_0 k_B T_0}\right)^{-\kappa_0-1}.
    \label{eq:n-phi}
\end{equation}

According to Eqs. \eqref{eq:fe-calc} and \eqref{eq:fe-calc-dev}, the electron distribution can be worked out by replacing $n_e(\phi)$ with Eq. \eqref{eq:n-phi},
\begin{equation}
    f_e(W) = n_0 \left( \frac{m}{2\pi k_B T_0 \kappa_0}\right)^{\frac{d}{2}} \frac{\Gamma(\kappa_0+1+\frac{d}{2})}{\Gamma(\kappa_0+1)} 
    \left( 1+ \frac{W}{\kappa_0 k_B T_0}\right)^{-\kappa_0-1-\frac{d}{2}}.
    \label{eq:fe}
\end{equation}
The detailed derivations can be found in the supplementary material.
The distribution \eqref{eq:fe} is the kappa distribution expressed by the invariant kappa index $\kappa_0$. \cite{Livadiotis2011a}
The constraint $\kappa_0>0$ is required to ensure a finite temperature. \cite{Livadiotis2010a}
With the transformation $\kappa = \kappa_0 + d/2$ and $d=3$, Eq. \eqref{eq:fe} recovers the 3-D kappa distribution widely used in the literature, \cite{Stverak2008,Stverak2009,Livadiotis2010a,Pierrard2016}
\begin{equation}
    f_e(W) = n_0 \left[\frac{m}{2 \pi k_B T_0 (\kappa - \frac{3}{2})} \right]^{\frac{3}{2}} \frac{\Gamma(\kappa+1)}{\Gamma(\kappa-\frac{1}{2})}
    \left(
    1 + \frac{1}{\kappa -\frac{3}{2}} \frac{W}{2 k_B T_0}
    \right)^{-\kappa-1}.
    \label{eq:fe-standard-kappa}
\end{equation}
It must be emphasized that the parameters $T_0$ in both Eqs. \eqref{eq:fe} and \eqref{eq:fe-standard-kappa} are the same because the transformation is only related to $\kappa_0$ and $\kappa$.
In addition, one can also prove that the kappa distribution \eqref{eq:fe} is the same as that derived in our previous work \cite{Guo2020}. The proof is shown in the supplementary material.

It is worth noting the differences between Eq. \eqref{eq:fe-standard-kappa} and another common form of the 3-D velocity kappa distribution, \cite{Vocks2008,Lazar2010,Cranmer2014}
\begin{equation}
   f_\kappa = \frac{n_0}{(\kappa \pi \theta^2)^{3/2}} \frac{\Gamma(\kappa+1)}{\Gamma(\kappa-\frac{1}{2})}
    \left( 1+ \frac{v^2}{\kappa \theta^2}\right)^{-\kappa-1},
    \label{eq:fe-alter-kappa}
\end{equation}
where $\theta$ is the most probable speed. 
The parameterizations of Eqs. \eqref{eq:fe-standard-kappa} and \eqref{eq:fe-alter-kappa} are different.
The independent parameters implied by Eq. \eqref{eq:fe-standard-kappa} are $\kappa$ and $T_0$, while those implied by Eq. \eqref{eq:fe-alter-kappa} are $\kappa$ and $\theta$.
Nevertheless, both of them have been used in the observations and theoretical works.
The kappa distribution \eqref{eq:fe-standard-kappa} has been used to fit the observation data of nonthermal electrons in the solar wind.\cite{Stverak2008,Stverak2009} 
Such a kappa distribution could provide a suitable description of the heat conduction in the solar corona. \cite{Landi2001}
Another mechanism, explaining the formation of the kappa distribution according to the weak turbulence theory, implies the temperature is independent of the kappa index. \cite{Yoon2014}
From the view of nonextensive statistical mechanics, Livadiotis also shows the independence of the kinetic temperature. \cite{Livadiotis2015a}
Besides, another kappa distribution \eqref{eq:fe-alter-kappa} has been applied to numerous studies. \cite{Maksimovic1997,Vocks2008,Dialynas2009,Lazar2010,Cranmer2014}
According to Ref. \onlinecite{Lazar2016}, both of them could be valid but for different plasma systems.

From the above derivations, one can conclude that the sufficient conditions to derive the kappa distribution \eqref{eq:fe} from the steady-state Vlasov equaiton \eqref{eq:V} are the polytropic EOS \eqref{assump:poly-eos} and Eq. \eqref{assump:w-func} .
Eq. \eqref{assump:w-func}, a valid result in one dimension but an assumption in higher dimensions, plays a very significant role in the present derivation.
On the one hand, this equation ensures that the expression of the distribution \eqref{eq:fe-calc} can be obtained from the definition of the electron density \eqref{def:ne}.
On the other hand, it also results in the zero fluid velocity and thus the pressure balance \eqref{eq:p-balance} in the stationary state.
As a result of the polytropic EOS \eqref{assump:poly-eos} and the pressure balance \eqref{eq:p-balance}, the electron density could be expressed as a function of the potential energy, namely Eq. \eqref{eq:n-phi}.
Combining Eqs. \eqref{eq:n-phi} and \eqref{eq:fe-calc}, one derives the kappa distribution \eqref{eq:fe}.

The present approach could be regarded as an improvement to our previous study \cite{Guo2020}.
First, the condition generating the kappa distribution in 1-D non-uniform plasma has been simplified.
Although the argument eliminating the assumption \eqref{assump:w-func} in 1 dimension is quite simple, the conclusion is not trivial.
Second, in higher dimensions, one of the assumptions in the previous work \cite{Guo2020}, i.e., Eq. \eqref{assump:previous2}, has been replaced with Eq. \eqref{assump:w-func}.
It has to be emphasized that these two assumptions are different, and there is no inclusion relation between them.
The functions satisfying the assumption \eqref{assump:w-func} must be a steady-state solution of the Vlasov equation, while those satisfying \eqref{assump:previous2} does not have such a property.
It means some of the functions satisfying \eqref{assump:previous2} may be steady-state solutions, but some others may not.
The new assumption \eqref{assump:w-func} has an advantage over \eqref{assump:previous2} because it is easier to be eliminated in some specific physical system where the energy is the only conservations, just like the 1-D case. 
However, one cannot eliminate the assumption \eqref{assump:previous2} by the same analysis.

\textit{The arbitrariness of the potential.} If the potential $\varphi$ is given, the solution of the Vlasov-Poisson system \eqref{eq:fe} can be totally determined. 
The density and temperature can also be determined from Eq. \eqref{eq:n-phi} and Eq. \eqref{dev:nphi2} respectively.
We will investigate two cases to discuss the constraints on the potential in this section.

In the first case, we consider a single species plasma only consisting of electrons, i.e., $n_i = 0$.
Substituting Eq. \eqref{eq:n-phi} into the Poisson equation \eqref{eq:P}, one finds,
\begin{equation}
    \nabla^2 \varphi = \frac{en_0}{\varepsilon_0} \left( 1-\frac{e\varphi}{\kappa_0 k_B T_0}\right)^{-\kappa_0-1}.
    \label{eq:le-eq-phi}
\end{equation} 
This equation could be transformed into the well-known Lane-Emden equation, \cite{Horedt2004,Binney2008}
\begin{equation}
    \nabla^2 \theta(\vb{r}^*) = -\theta^{-\kappa_0-1},
    \label{eq:le-eq}
\end{equation}
by denoting
\begin{equation}
    \theta = 1-\frac{e\varphi}{\kappa_0 k_B T_0}, \quad 
    \vb{r}^* = \sqrt{\frac{n_0 e^2}{\kappa_0 \varepsilon_0 k_B T_0}} \vb{r}.
\end{equation}
It indicates that the self-consistent potential must be a solution of Eq. \eqref{eq:le-eq} under some boundary conditions.
Therefore, the potential cannot be an arbitrary function in the case of a single species plasma.
The kappa index and the boundary conditions directly determine the profile of the potential through the Lane-Emden equation \eqref{eq:le-eq}.

In the second case, a two-component plasma is considered, i.e., $n_i \neq 0$.
In this case, if we may construct the ion density in the following manner,
\begin{equation}
    n_i = n_e - \frac{\varepsilon_0}{e} \nabla^2 \varphi,
    \label{eq:n_i}
\end{equation}
then the electron density, temperature, and distribution function can all exist with the arbitrary potential $\varphi$ without breaking the self-consistency in terms of Eqs. \eqref{dev:nphi2}-\eqref{eq:fe}.

\textit{The connections with the polytrope.} The polytrope is an old and well-known model that describes the self-gravitating system following the polytropic EOS.
The 3-D distribution of the polytrope is, \cite{Binney2008}
\begin{equation}
    f(W) = \left\{
    \begin{aligned}
        & F (\Phi_0-W)^{\nu - \frac{3}{2}} \quad & (W<\Phi_0, \nu > \frac{1}{2}), \\
        & 0 \quad & (W \geq \Phi_0),
    \end{aligned}
    \right.
    \label{eq:polytrope}
\end{equation}
where $\Phi_0$ is some constant, $\nu$ is an index related to the polytropic index by
\begin{equation}
    \gamma = 1 + \frac{1}{\nu},
\end{equation}
and $F$ is the normalization. 
If we let 
\begin{equation}
    \Phi_0 = -\kappa_0 k_B T_0 \quad \text{and} \quad \nu = -\kappa_0-1,
\end{equation}
then we find the polytrope distribution \eqref{eq:polytrope}
\begin{equation}
    f(W) \propto \left( 1 - \frac{W}{-\kappa_0 k_B T_0} \right)^{-\kappa_0-1-\frac{3}{2}},
\end{equation}
which is a similar mathematical form as the kappa distribution \eqref{eq:fe} with $d=3$.

However, we have to stress that the polytrope model and the kappa model are different.
The reasons are as follows.
First, the value ranges of the polytropic indices are different. 
For the polytrope, the constraint that $\nu > \frac{1}{2}$ is equivalent to $\kappa_0<-\frac{3}{2}$ and $1<\gamma<3$.
But for the kappa distribution, $\kappa_0>0$ is required to ensure a finite temperature, which indicates that the polytropic index must satisfy $0<\gamma<1$, or equivalently $\nu<-1$.
The polytrope distribution with the constraint $\nu>\frac{1}{2}$ may be regarded as a "negative" kappa distribution with $\kappa_0 < - \frac{3}{2}$.
Second, the systems described by these two models are not the same. 
The polytrope model always describes the self-gravitating system consisting of only one component.
However, the kappa model could describe not only one-component but also multi-component plasma systems. \cite{Baluku2015,Hatami2018}

\textit{Summary.} In this study, we investigate the sufficient conditions to derive the kappa distribution \eqref{eq:fe} in the inhomogeneous isotropic Vlasov-Poisson system by an improved method compared with our previous study \cite{Guo2020}. 
In one dimension, only the polytropic assumption \eqref{assump:poly-eos} is needed in the present work.
However, in higher dimensions, we still need an additional assumption that the global distribution must only be the function of energy. 
We also prove that the kappa distribution \eqref{eq:fe} is equivalent to that obtained in Ref. \onlinecite{Guo2020}.
Besides, the constraints on the potential are discussed.
It shows that the self-consistent potential is governed by the Lane-Emden equation \eqref{eq:le-eq} in a one-component plasma.
In the case of a two-component plasma, any potential could be self-consistent if we construct the ion density by Eq. \eqref{eq:n_i}.
At last, the connections between the kappa distribution and the polytrope are discussed.

\section*{Supplementary Material}
See the supplementary material for (1) the calculations of the distribution function; and (2) the equivalence between the kappa distributions.

\begin{acknowledgments}
This work was supported by the National Natural Science Foundation of China (No.11775156) and by the Fundamental Research Funds for the Central Universities, Civil Aviation University of China (No.3122019138).
\end{acknowledgments}

\section*{Data Availability}
Data sharing is not applicable to this article as no new data were created or analyzed in this study.

\bibliography{mylib}

\newpage
\renewcommand{\theequation}{S\arabic{equation}}
\setcounter{equation}{0}
\centerline{\Large Supplementary Material} 
\vspace{0.5cm}
\section{Calculations of the distribution function}
\label{ap:calc-fe}
With $n_e(\phi)$ given by Eq. (21), the distribution $f_e(W)$ could be solved from Eqs. (12) and (13) in the main text.
When $d$ is even, $f_e$ can be calculated directly,
\begin{align}
    f_e(W) &= n_0 \left(\frac{m}{2\pi}\right)^{\frac{d}{2}} \left( -\dv{}{W}\right)^{\frac{d}{2}} 
             \left( 1 + \frac{W}{\kappa_0 k_B T_0}\right)^{-\kappa_0-1} \notag \\
           &= n_0 \left( \frac{m}{2\pi k_B T_0 \kappa_0}\right)^{\frac{d}{2}} \frac{\Gamma(\kappa_0+1+\frac{d}{2})}{\Gamma(\kappa_0+1)} \left( 1+ \frac{W}{\kappa_0 k_B T_0}\right)^{-\kappa_0-1-\frac{d}{2}}.
           \label{ap-eq:fe-even}
\end{align}
When $d$ is odd, $f_e$ is,
\begin{equation}
    f_e(W) = \frac{n_0}{\Gamma(\frac{1}{2})} \left(\frac{m}{2\pi}\right)^{\frac{d}{2}} \left( -\dv{}{W}\right)^{\frac{d+1}{2}} I, 
    \label{ap-eq:calc-fe-odd}
\end{equation}
with
\begin{equation}
    I = \int_W^\infty(\phi-W)^{-\frac{1}{2}} \left( 1+\frac{\phi}{\kappa_0 k_B T_0}\right)^{-\kappa_0-1} \dd{\phi}.
\end{equation}
Let $\phi^* = \phi-W$, the integral $I$ turns out to be,
\begin{align}
    I &= \int_0^\infty {\phi^*}^{-\frac{1}{2}}
        \left(1+\frac{\phi^*+W}{\kappa_0 k_B T_0}\right)^{-\kappa_0-1} \dd{\phi^*} \notag \\
      &= \left(1+\frac{W}{\kappa_0 k_B T_0}\right)^{-\kappa_0-1} 
        \int_0^\infty {\phi^*}^{-\frac{1}{2}} 
            \left(1+\frac{\phi^*}{W+\kappa_0 k_B T_0}\right)^{-\kappa_0-1} \dd{\phi^*} \notag \\
      &= (\kappa_0 k_B T_0)^{\frac{1}{2}} \left(1+\frac{W}{\kappa_0 k_B T_0}\right)^{-\kappa_0-1} 
         \frac{\Gamma(\frac{1}{2}) \Gamma(\kappa_0+\frac{1}{2})}{\Gamma(\kappa_0+1)},
    \label{ap-eq:integral}
\end{align}
where the formula of the beta integral, \cite{Olver2010}
$\Gamma(a)\Gamma(b) / \Gamma(a+b) = \int_0^\infty t^{a-1}(1+t)^{-a-b} \dd{t},$
is used.
Substituting Eq. \eqref{ap-eq:integral} into Eq. \eqref{ap-eq:calc-fe-odd}, we derive the distribution,
\begin{align}
    f_e(W) &= n_0 (\kappa_0 k_B T_0)^{\frac{1}{2}}
                \frac{\Gamma(\kappa_0+\frac{1}{2})}{\Gamma(\kappa_0+1)}
                \left(\frac{m}{2\pi}\right)^{\frac{d}{2}} 
                \left( -\dv{}{W}\right)^{\frac{d+1}{2}}  \left(1+\frac{W}{\kappa_0 k_B T_0}\right)^{-\kappa_0-1} \notag \\
           &= n_0 \left( \frac{m}{2\pi k_B T_0 \kappa_0}\right)^{\frac{d}{2}} \frac{\Gamma(\kappa_0+1+\frac{d}{2})}{\Gamma(\kappa_0+1)} \left( 1+ \frac{W}{\kappa_0 k_B T_0}\right)^{-\kappa_0-1-\frac{d}{2}},
\end{align}
which is the same as the result for an even $d$, i.e., Eq. \eqref{ap-eq:fe-even}.

\section{Equivalence between the kappa distributions}
\label{ap:equivalence}
The kappa distribution derived in our previous study reads, \cite{Guo2020}
\begin{equation}
    f_{e} (\vb{r},\vb{v}) = n_e(\vb{r}) \left[ \frac{m}{2 \pi k_B T(\vb{r}) \kappa_0} \right]^ \frac{d}{2}  \frac{\Gamma(\kappa_0+1+ \frac{d}{2})}{\Gamma(\kappa_0+1)}
    \left[ 1+ \frac{1}{\kappa_0} \frac{m v^2}{2k_BT(\vb{r})} \right ]^{ -\kappa_0 - 1 - \frac{d}{2} },
    \label{ap-eq:fe-pre}
\end{equation}
where the local temperature $T(\vb{r})$ can be expressed as,
\begin{equation}
    T(\vb{r}) = T_0 \left[\frac{n_e(\vb{r})}{n_0}\right]^{- \frac{1}{\kappa_0+1} },
    \label{ap-eq:T-pre}
\end{equation}
and the potential
\begin{equation}
    \varphi(\vb{r}) = \varphi_0 - \kappa_0 \frac{k_B T_0}{e} \left[\frac{n_e(\vb{r})}{n_0}\right]^{- \frac{1}{\kappa_0+1} }.
    \label{ap-eq:phi-pre}
\end{equation}
Because we choose the zero potential point at the position $T = T_0$, we have $\varphi_0 = \kappa_0 k_B T_0$.
Then, in terms of Eq. \eqref{ap-eq:phi-pre}, we can express the density as,
\begin{equation}
    n_e = n_0 \left( 1 - \frac{e\varphi}{\kappa_0 k_B T_0}   \right)^{-\kappa_0-1},
    \label{ap-eq:n_e}
\end{equation}
as well as the temperature,
\begin{equation}
    T = T_0 \left( 1 - \frac{e\varphi}{\kappa_0 k_B T_0}\right).
    \label{ap-eq:T}
\end{equation}
Substituting the density \eqref{ap-eq:n_e} and the temperature \eqref{ap-eq:T} into Eq. \eqref{ap-eq:fe-pre}, one finds,
\begin{align}
    f_e =& n_0 \left(1-\frac{e\varphi}{\kappa_0 k_B T_0}\right)^{-\kappa_0-1}
            \left[ \frac{m}{2\pi (\kappa_0 k_B T_0 - e\varphi)}  \right]^{ \frac{d}{2} }
            \frac{\Gamma(\kappa_0+1+\frac{d}{2})}{\Gamma(\kappa_0+1)} 
            \left( 1+ \frac{\frac{1}{2}mv^2}{\kappa_0 k_B T_0 - e\varphi} \right)^{-\kappa_0-1-\frac{d}{2} } \notag \\
        =& n_0 (\kappa_0 k_B T_0)^{\kappa_0+1} \left(\frac{m}{2\pi} \right)^{\frac{d}{2}}
            \frac{\Gamma(\kappa_0+1+\frac{d}{2})}{\Gamma(\kappa_0+1)} 
            \left( \kappa_0 k_B T_0 + \frac{1}{2}mv^2 - e\varphi \right)^{-\kappa_0-1-\frac{d}{2}} \notag \\
        =& n_0 \left( \frac{m}{2\pi k_B T_0 \kappa_0}  \right)^{\frac{d}{2}}
            \frac{\Gamma(\kappa_0+1+\frac{d}{2})}{\Gamma(\kappa_0+1)} 
            \left( 1+ \frac{\frac{1}{2}mv^2-e\varphi}{\kappa_0 k_B T_0} \right)^{-\kappa_0-1-\frac{d}{2} },
\end{align}
which is the same as Eq. (22) in the main text.
\end{document}